\documentclass[aps,prb,reprint,superscriptaddress,longbibliography]{revtex4-2}

\usepackage[colorlinks,linkcolor=blue,citecolor=blue,urlcolor=blue]{hyperref}

\usepackage{amsmath,amssymb,amsfonts,bm,graphicx}
\usepackage{microtype}
\usepackage{comment}

\newcommand{\CP}{\mathbb{CP}^{2}}
\newcommand{\SU}{\mathrm{SU}}

\newcommand{\dd}{\mathrm{d}}
\newcommand{\BZ}{\mathrm{BZ}}
\newcommand{\Fig}[1]{Fig.~\ref{#1}}
\usepackage{comment}

\begin{document}
\title{Non-Abelian Hall Effects of Flavor-Waves in $\mathbb{CP}^{2}$ Skyrmions}

\author{Carlos Saji}
\email{carlos.saji@usach.cl}
\affiliation{Departamento de Física, Universidad de Santiago de Chile, 9170124 Santiago, Chile.}
\author{Eduardo Saavedra}
\email{eduardo.saavedra.d@usach.cl}
\affiliation{Departamento de Física, Universidad de Santiago de Chile, 9170124 Santiago, Chile.}

\author{Alvaro S. Nunez}
\affiliation{Departamento de F\'isica, CEDENNA, FCFM, Universidad de Chile, Santiago, Chile}
\author{Roberto E. Troncoso}
\affiliation{Instituto de Alta Investigaci\'on, Universidad de Tarapac\'a, Casilla 7D, Arica, Chile}

\begin{abstract}
We investigate the collective excitations of isolated and crystalline $\CP$ skyrmions in a two-dimensional triangular-lattice spin-1 magnet. Within an $\SU(3)$ coherent-state framework, fluctuations of the magnetic dipole and quadrupole moments are described by two coupled bosonic flavors. In a uniform spin-polarized state, these flavors can be identified with single-spin-flip and double-spin-flip excitations, whereas a nonuniform $\CP$ texture mixes their character. We formulate the corresponding bosonic Bogoliubov--de Gennes problem and examine the matrix-valued gauge connection governing flavor-wave propagation. For isolated skyrmions, the calculated spectrum contains localized modes, and the scattering dynamics exhibit transverse deflection accompanied by isospin precession. For a numerically relaxed skyrmion crystal, we obtain collective-mode bands with nonzero Chern numbers and calculate the associated thermal Hall response. These findings illustrate how projective magnetic textures couple multipolar excitations to non-Abelian transport and topological band phenomena.
\end{abstract}
\maketitle

\section{Introduction}\label{sec:intro}

Collective magnetic excitations are conventionally formulated as spin waves, describing fluctuations transverse to an ordered dipole moment. This picture becomes incomplete in magnets with local spin $S\geq 1$, for which higher-rank magnetic multipoles constitute additional dynamical degrees of freedom. In particular, spin-1 moments possess both dipolar and quadrupolar components, and interactions involving the latter can stabilize spin-nematic order even when the expectation value of the magnetic dipole vanishes \cite{Papanicolaou1988,Luchli2006,Tsunetsugu2006,Penc2010}. To describe such excitations, one must to extend the conventional spin wave model. This extension leads to the linear flavor-wave theory (LFWT) developed by Papanicolaou \cite{Papanicolaou1984,Papanicolaou1988}. Overall, this formalism has been used to calculate the linear fluctuations spectrum about the ferroquadrupolar and nematic ground states for spin-1 magnets with bilinear-biquadratic (BBQ) energy model \cite{Muniz2014,Remund2022,Smerald2013,Luchli2006,Seifert2022,Pires2017}, as well as, for frustrated antiferromagnets with $SU(N)$ symmetry \cite{Jin2019,Pohle2024,Natori2019,Kim2017}. 
Furthermore, in recent works, quadrupole waves have been observed experimentally \cite{Sheng2025,Watanabe2025,Nag2022,Kim2023} , where hybridized dipole-quadrupolar moment waves have been reported \cite{Bai2021}, in agreement with theorical results obtained from LFWT.

On the other hand, the existence and stability of topologically nontrivial textures in spin-1 quantum magnets, namely $\CP$ skyrmions, have been addressed in exhautuvely in several works \cite{Akagi2021_2,Akagi2021,Amari2022,Zhang2023,Eto2013,Garaud2013,Benfenati2023,Benfenati2023,Azaria1995,Amari2026,Rajaraman2002,Ferreira2025}. Moreover, recently it is has proved that $\CP$ skyrmions exhibit collective dynamics through the equation of Thiele and also topological Hall effect of electrons \cite{Lee2025}. 

Despite the growing interest in $\CP$ skyrmions and the substantial progress in understanding their topological properties, stability, and static configurations, their dynamical response to collective excitations remains largely unexplored. In particular, the interaction between nontrivial $\CP$ magnetic textures and propagating flavor waves has received comparatively little attention. Unlike conventional spin waves in dipolar magnets, flavor waves in spin-1 systems involve coupled dipolar and quadrupolar degrees of freedom, providing additional channels through which the underlying topology can influence quasiparticle dynamics. This raises fundamental questions regarding the emergence of effective gauge fields, the hybridization of excitation modes, and the possibility of unconventional transverse transport phenomena. Addressing these questions is essential for establishing a comprehensive understanding of the dynamical and transport properties of $\CP$ skyrmion textures. In this work, we investigate these effects within a microscopic flavor-wave framework, revealing how the nontrivial geometry of $\CP$ skyrmions gives rise to topological flavor bands and associated flavor Hall effect.

The propagation of collective excitations through such textures raises questions that do not arise in ordinary single-flavor magnonics. In particular, the local tangent basis varies with position and can induce a matrix-valued connection acting on the two-component flavor-wave amplitude. Non-Abelian approaches have already proved useful for describing coupled spin-wave polarizations in antiferromagnets \cite{Daniels2018,Daniels2019}; related gauge structures have been implicated in the magnon thermal Hall effect of antiferromagnetic skyrmion lattices \cite{Takeda2024}. The $\CP$ problem differs in the microscopic origin of its internal degrees of freedom: the two flavors describe fluctuations of a single local spin-1 state, rather than being introduced solely through distinct magnetic sublattices. This setting provides an opportunity to examine flavor conversion, isospin dynamics, and transverse wave propagation within a common multipolar framework.

A separate issue concerns the momentum-space topology of collective excitations in a periodic projective texture. The real-space topological charge of a skyrmion does not determine the Chern number of an excitation band by itself. Establishing topological flavor-wave transport therefore requires an explicit calculation of the bosonic spectrum, its Berry curvature, and the response obtained from the thermally occupied modes. Understanding these connections is especially relevant in spin-1 systems, where hybridization can redistribute dipolar and quadrupolar character across the excitation bands.

In this work, we investigate these questions in a triangular-lattice spin-1 model with bilinear-biquadratic exchange, Dzyaloshinskii--Moriya (DM) interaction, magnetic anisotropy, and an external field. After characterizing the $\CP$ order parameter and numerically relaxed textures, we formulate a two-flavor theory of small oscillations and identify the relevant spin and quadrupole sectors in the polarized reference state. We then examine the non-Abelian gauge description of flavor-wave propagation, localized modes and scattering by an isolated skyrmion, and the topological bands and thermal Hall response of a skyrmion crystal. Our analysis focuses on the numerical results and analytical expressions obtained for the model considered here.

\section{Model}\label{sec:model}

We consider a spin-1 system on a two-dimensional triangular lattice with lattice constant $a$. We introduce the five-component vector $\bm Q_j$ by the quadrupolar degrees of freedom,
\begin{equation*}
\bm Q_j=\begin{pmatrix}
Q^{x^2-y^2}\\ Q^{r^2-3z^2}\\Q^{xy}\\Q^{yz}\\Q^{xz}
\end{pmatrix}_{\!j}
=\begin{pmatrix}
(S^x)^2-(S^y)^2\\
\frac{1}{\sqrt3}\left(\bm S^2-3(S^z)^2\right)\\
S^xS^y+S^yS^x\\S^yS^z+S^zS^y\\S^xS^z+S^zS^x
\end{pmatrix}_{\!j}.
\end{equation*}
The quadrupole--quadrupole coupling can equivalently be expressed as $\bm Q_i\cdot\bm Q_j=\frac12\sum_{\alpha\beta}Q_i^{\alpha\beta}Q_j^{\alpha\beta}$, where $Q^{\alpha\beta}=S^\alpha S^\beta+S^\beta S^\alpha-\frac{2}{3}S(S+1)\delta^{\alpha\beta}$ denotes the quadrupolar tensor. 

To treat dipolar and quadrupolar degrees of freedom on equal footing, we introduce the eigenstates $\{|1\rangle_j,|0\rangle_j,|-1\rangle_j\}$ of $\hat S_j^z$ and the time-reversal-invariant basis $|x^1\rangle_j=\frac{i}{2}(|1\rangle_j-|-1\rangle_j)$, $|x^2\rangle_j=\frac12(|1\rangle_j+|-1\rangle_j)$, and $|x^3\rangle_j=-i|0\rangle_j$. A product coherent state is specified by $|Z\rangle=\bigotimes_j|Z_j\rangle$, with $|Z_j\rangle=\sum_{\alpha=0}^{3}Z_j^\alpha|x^\alpha\rangle_j$ and $Z_j=(Z_j^1,Z_j^2,Z_j^3)^T\in\mathbb C^3$, subject to $\langle Z_j|Z_j\rangle=1$. The local order parameter is determined by the expectation values
\begin{equation*}
n_j^\mu=\langle Z_j|\hat\lambda^\mu|Z_j\rangle,
\end{equation*}
where $\hat\lambda^\mu$ ($\mu=1,\ldots,8$) are the Gell-Mann generators of $\mathfrak{su}(3)$. Their normalization and algebra are
\begin{align*}
\operatorname{Tr}(\hat\lambda_\mu\hat\lambda_\nu)&=2\delta_{\mu\nu},\\
[\hat\lambda_\mu,\hat\lambda_\nu]&=2i f_{\mu\nu\eta}\hat\lambda_\eta,\\
\{\hat\lambda_\mu,\hat\lambda_\nu\}&=\frac{4}{3}\delta_{\mu\nu}\hat I+2d_{\mu\nu\eta}\hat\lambda_\eta.
\end{align*}
Here and below repeated $SU(3)$ indices are summed. The vector $\bm n_i\in\mathbb R^8$ encodes the three spin and five quadrupole expectations.In this basis the spin and quadrupole operators read $\hat{\bm S}=(\hat\lambda^7,-\hat\lambda^5,\hat\lambda^2)^T$ and $\hat{\bm Q}=-(\hat\lambda^3,\hat\lambda^8,\hat\lambda^1,\hat\lambda^4,\hat\lambda^6)^T.$ We recall that the eight real coordinates are not independent: 
\begin{equation}
 \sum_{\mu=1}^{8}(n_i^\mu)^2=\frac43,
 \qquad n_i^\mu=\frac32d_{\mu\nu\rho}n_i^\nu n_i^\rho,
\label{eq:pureconstraints}
\end{equation}
where $d_{\mu\nu\rho}$ is the symmetric $SU(3)$ tensor defined through the anticommutator of the generators. These constraints select the four-real-dimensional manifold $\CP$ inside $\mathbb R^8$.

\subsection{Hamiltonian and coherent-state variables}

We consider spin-one moments on a triangular lattice of lattice constant $a$. Our starting Hamiltonian contains bilinear and quadrupolar exchange, Dzyaloshinskii--Moriya (DM) interactions, a Zeeman field. the The isotropic tight-binding $SU(3)$ coherent-state energy is given by
\begin{equation}
E[\bm n]=-\sum_{\langle ij\rangle}\left(J\bm n_i\cdot\bm n_j+
 f_{\mu\nu\eta}D_{ij}^{\mu}n_i^{\nu}n_j^{\eta}\right)
 -h\sum_i  n_{2},
\label{eq:energy_n}
\end{equation}
where $h=g\mu_B B$. The DM term in Eq.~\eqref{eq:energy_n} represents the fully $SU(3)$-symmetric generalized DM interaction. We use a bond-dependent in-plane DM vector $\bm D_{ij}=D (\bm{r}_{i}-\bm{r}_{j})\cdot \hat{\bm{S}}$ consistent with the triangular-lattice geometry. The first term in \ref{eq:energy_n} corresponds to the bilinear-biquadratic (BBQ) exchange energy, which allows magnetic and nematic ordering tendencies to compete \cite{Luchli2006,Wenzel2013,Niesen2018,Seifert2022}. 

The semiclassical dynamics of spin-1 magnets is governed by the generalized Landau--Lifshitz--Gilbert equation \cite{Dahlbom2022,Remund2022}
\begin{equation}
\frac{\partial n_j^\mu}{\partial t}
+\alpha f_{\mu\nu\eta}n_j^\nu\frac{\partial n_j^\eta}{\partial t}
=\gamma_{0} f_{\mu\nu\eta}n_j^\nu\frac{\partial E}{\partial n_j^\eta},
\label{eq:llg}
\end{equation}
where the terms with $\gamma_{0}$ and $\alpha$ describe conservative and dissipative contributions, respectively.

The topology of a two-dimensional $\CP$ texture follows from the second homotopy group $\pi_2(\CP)=\mathbb Z$. The topological charge is calculated by the integral \cite{Amari2022,Zhang2023,Akagi2021_2}
\begin{equation}
\mathcal{Q}_{\mathrm{top}}=\frac{1}{8\pi}\int\epsilon^{jk}f_{\mu\nu\eta}
 n^\mu\partial_jn^\nu\partial_kn^\eta\,\dd^2x\in\mathbb Z.
\label{eq:charge}
\end{equation}
For lattice configurations, we evaluate this quantity through the geodesic-interpolation construction $\mathcal{Q}_{\mathrm{top}}=\sum_{\Delta_{jkl}}\rho_{jkl}=\frac{1}{2\pi}\sum_{\Delta_{jkl}}(\gamma_{jl}+\gamma_{lk}+\gamma_{kj})$, where the sum runs over oriented triangular plaquettes and $\gamma_{jk}=\arg\langle Z_j|Z_k\rangle$. This prescription connects the discrete spinor texture to the continuum definition in Eq.~\eqref{eq:charge}.

\begin{figure}[t]
\centering\includegraphics[width=\columnwidth]{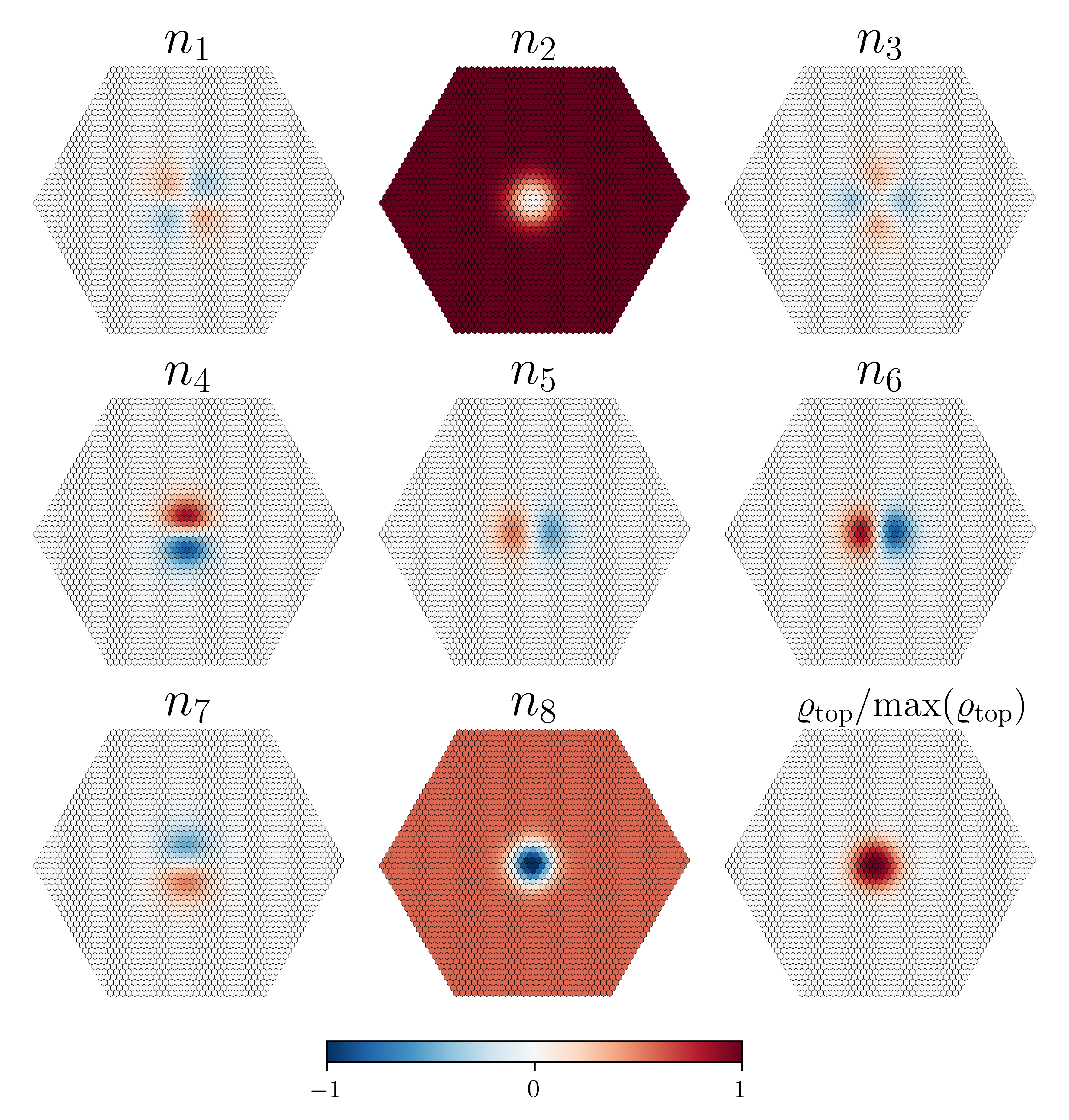}
\caption{Stabilized isolated $\CP$ skyrmion on a triangular lattice with open boundary conditions, for parameters: $J=1\,\mathrm{mV}$, $D=0.3J$, and $h=0.7J$. The calculated topological charge yields $\mathcal{Q}_{\mathrm{top}}\approx1$.}
\label{fig:isolated}
\end{figure}

The static texture is obtained by minimizing the energy with the normalization condition $\langle Z_j|Z_j\rangle=1$. The corresponding stationary equations are
\begin{align*}
\frac{\partial n_{j}^\mu}{\partial\operatorname{Re}Z^\alpha}
\frac{\delta E}{\delta n_{j}^\mu}-c\operatorname{Re}Z_{j}^\alpha&=0,\\
\frac{\partial n_{j}^\mu}{\partial\operatorname{Im}Z_{j}^\alpha}
\frac{\delta E}{\delta n_{j}^\mu}-c\operatorname{Im}Z_{j}^\alpha&=0,
\end{align*}
where $c$ is the Lagrange multiplier. These equations are solved numerically using a nonlinear conjugate-gradient procedure. A stabilized isolated $\CP$ skyrmion is shown in \Fig{fig:isolated}: the components of $\bm n$ reveal the spatial organization of both dipolar and quadrupolar order.

Note that, in contrast to conventional skyrmions described by a unit vector on $S^2$, projective skyrmions can involve simultaneous spatial variations of dipolar and quadrupolar order, as illustrates Fig. \ref{fig:isolated}.

\subsection{Linear flavor-wave theory in a $\CP$ textures}\label{sec:flavor}

Then, following \cite{Remund2022}, we introduce the pair of $SU(3)$ Schwinger bosons (SBs) $\psi_{1},\psi_{2}$, where $\psi_{1}^{\dagger}$ corresponds to creating a 1-magnon dipolar excitation, whereas $\psi_{2}^{\dagger}$ corresponds to creating a local  2-magnon quadrupolar excitation.

We now determine the spectrum of the fluctuations about an equilibrium state $\bm n_{0}=\langle Z_{0}|\hat{\bm\lambda}|Z_{0}\rangle$. For a magnetic unit cell containing $N$ sites, we introduce $\hat{\bm{\psi}}_{\bm{k}}=(\psi^{1}_{\bm{k}},\psi^{2}_{\bm{k}})$ with $2N$ components and the Nambu spinor $\hat{\Psi}_{\bm k}=(\hat{\bm{\psi}}_{\bm k},\hat{\bm{\psi}}_{-\bm k}^\dagger)^T$. We can write de Hamiltonian of the linear fluctuations in the form $\hat{H}=\frac12\sum_{\bm k\in BZ} \hat{\Psi}_{\bm k}^{\dagger} \mathcal{H}_{\rm BdG}(\bm k)\hat{\Psi}_{\bm k}$, where the Bogoliubov–de Gennes (BdG) Hamiltonian reads (see the Supplemental Material for more details):
\begin{align}
 \mathcal H_{\rm BdG}(\bm k)&=
 \begin{pmatrix}
 \Xi(\bm k)& \Delta(\bm k)\\
 \Delta^{*}(-\bm k)& \Xi^{*}(-\bm k)
 \end{pmatrix},\label{eq:BdG}
\end{align}
The bosonic eigenvalue problem is not an ordinary Hermitian diagonalization, but rather
\begin{equation}
 \Sigma_z\mathcal H_{\rm BdG}(\bm k)\Psi_{n\bm k}
 =\varepsilon_{n\bm k}\Psi_{n\bm k}
\label{eq:geneig}
\end{equation}
where $[\Sigma_z]_{\beta j}^{\alpha i}=[\sigma_z]_{\alpha\beta}[I_{2N}]_{ij}$. We diagonalize the quadratic Hamiltonian with a paraunitary transformation $U$ following the Colpa procedure,
\begin{align*}
T_{\bm k}^\dagger \mathcal H_{\rm BdG}(\bm k) T_{\bm k}&=\mathbb{I}_2\otimes\operatorname{diag}(\varepsilon_0(\bm k),\varepsilon_1(\bm k),\ldots,\varepsilon_{2N-1}(\bm k)),
\end{align*}
and $T_{\bm k}^\dagger\Sigma_z T_{\bm k}=\Sigma_z.$ The quasiparticle operators $(\hat\gamma_{\bm k},\hat\gamma_{\bm k}^\dagger)^T=T_{\bm k}^{-1}(\hat\psi_{\bm k},\hat\psi_{-\bm k}^\dagger)^T$ then bring the Hamiltonian into the diagonal form
\begin{equation*}
\hat{H}=\sum_{\bm{k},n}\omega_n(\bm k)\left(\hat\gamma_{\bm{k},n}^\dagger\hat\gamma_{\bm{k},n}+\frac12\right).
\end{equation*}
The spatial amplitude of the $n$th mode is $\Psi_n(\bm k)_j=\langle0|\hat\gamma_n^\dagger\hat\psi_j|0\rangle=[T_{\bm k}]_{jn}$, with the magnon vacuum denoted by $|\mathrm{vac}\rangle=\prod_{j,n}\hat\gamma_n|0\rangle$. Thus, the calculation provides both the excitation energies and the spatial structure of the collective modes. We characterize their multipolar content through the spin and quadrupole fluctuation densities $(\delta\langle\bm S\rangle)^2=\Psi^\dagger\hat\varrho_S(\bm x)\Psi$ and $(\delta\langle\bm Q\rangle)^2=\Psi^\dagger\hat\varrho_Q(\bm x)\Psi$, with
\begin{equation*}
[\hat\varrho_S(\bm x)]_{\alpha\beta}=
\sum_{\mu=2,5,7}e_\alpha^\mu(\bm x)e_\alpha^\mu(\bm x),
\end{equation*}
and $[\hat\varrho_Q(\bm x)]_{\alpha\beta}=\delta_{\alpha\beta}-[\hat\varrho_S(\bm x)]_{\alpha\beta}$. These quantities provide a convenient way to discuss the relative dipolar and quadrupolar contributions in the calculated modes.

\subsection{Dipolar and quadrupolar character}\label{sec:ferro}

To illustrate the previous formalism, let us first consider a fully polarized state, $|\bm{Z}^{(0)}\rangle=\bigotimes_{j}|S_j^z=+1\rangle$, so that $\bm n^{\mathrm{FM}}=(0,1,0,0,0,0,0,1/\sqrt3)^T$. For this case, the independent Schwinger bosons read
\begin{equation*}
\hat\psi_{1,j}=\frac{\hat S_j^x+i\hat S_j^y}{\sqrt2},\qquad
\hat\psi_{2,j}=\frac{\hat Q_j^{x^2-y^2}+i\hat Q_j^{xy}}{2}.
\end{equation*}
Hence, we have that the change in the spin $z$-component turns out 
\begin{equation*}
\Delta\hat S_j^z=1-\hat S_j^z\approx
\hat\psi_{1,j}^\dagger\hat\psi_{1,j}
+2\hat\psi_{2,j}^\dagger\hat\psi_{2,j}.
\end{equation*}
 In consequence, the dipolar and quadrupolar moment excitations carry a quantized angular momentum of $l=1$ and $l=2$, respectively.  
 
 For the fully polarized state, the momentum-space Hamiltonian becomes
\begin{equation*}
\mathcal H(\bm k)=
\begin{pmatrix}
\hat h(\bm k)&0\\
0&\hat h^{*}(-\bm k)
\end{pmatrix},\ \hat h(\bm k)=\begin{pmatrix}
f(\bm k)+\frac{h}{2}&g^*(\bm k)\\
g(\bm k)&f(\bm k)+ h
\end{pmatrix}
\end{equation*}
where $f(\bm k)=-J\sum_{\bm\delta}e^{i\bm k\cdot\bm\delta}+zJ$ and $g(\bm k)=\frac{i D}{\sqrt2}\sum_{\bm\delta}e^{i(\bm k\cdot\bm\delta+\theta_{\bm\delta})}$, with $\bm\delta=\bm r_j-\bm r_i$ denoting the bond vectors connecting nearest
neighbors sites, $\theta_{\bm\delta}$ specifies their orientation relative to the $x$ axis, and $z=6$ is the coordination number. The corresponding diagonal form is
\begin{equation*}
\hat H=\sum_{\bm k}\left[\varepsilon^{(-)}(\bm k)
\hat\alpha_{\bm k}^\dagger\hat\alpha_{\bm k}
+\varepsilon^{(+)}(\bm k)\hat\beta_{\bm k}^\dagger\hat\beta_{\bm k}\right].
\end{equation*}
Here $\hat\psi_{1,\bm k}=u^{(-)}(\bm k)\hat\alpha_{\bm k}+v^{(-)}(\bm k)\hat\beta_{\bm k}$ and $\hat\psi_{2,\bm k}=u^{(+)}(\bm k)\hat\alpha_{\bm k}+v^{(+)}(\bm k)\hat\beta_{\bm k}$, with eigenvectors $(u^{(\pm)},v^{(\pm)})$ of $\hat h(\bm k)$. The two branches $\varepsilon^{(\pm)}(\bm k)$ describe the eigenstates $|\bm k,-\rangle=\hat\alpha^\dagger|\mathrm{vac}\rangle$ and $|\bm k,-\rangle=\hat\beta^\dagger|\mathrm{vac}\rangle$. For these eigenmodes the spin and quadrupolar densities are
\begin{align}
(\delta\langle\bm S\rangle)^2&=|u^{(\pm)}(\bm k)|^2,\\
(\delta\langle\bm Q\rangle)^2&=|u^{(\pm)}(\bm k)|^2+2|v^{(\pm)}(\bm k)|^2.
\end{align}

\begin{figure}[t]
\centering\includegraphics[width=0.88\columnwidth]{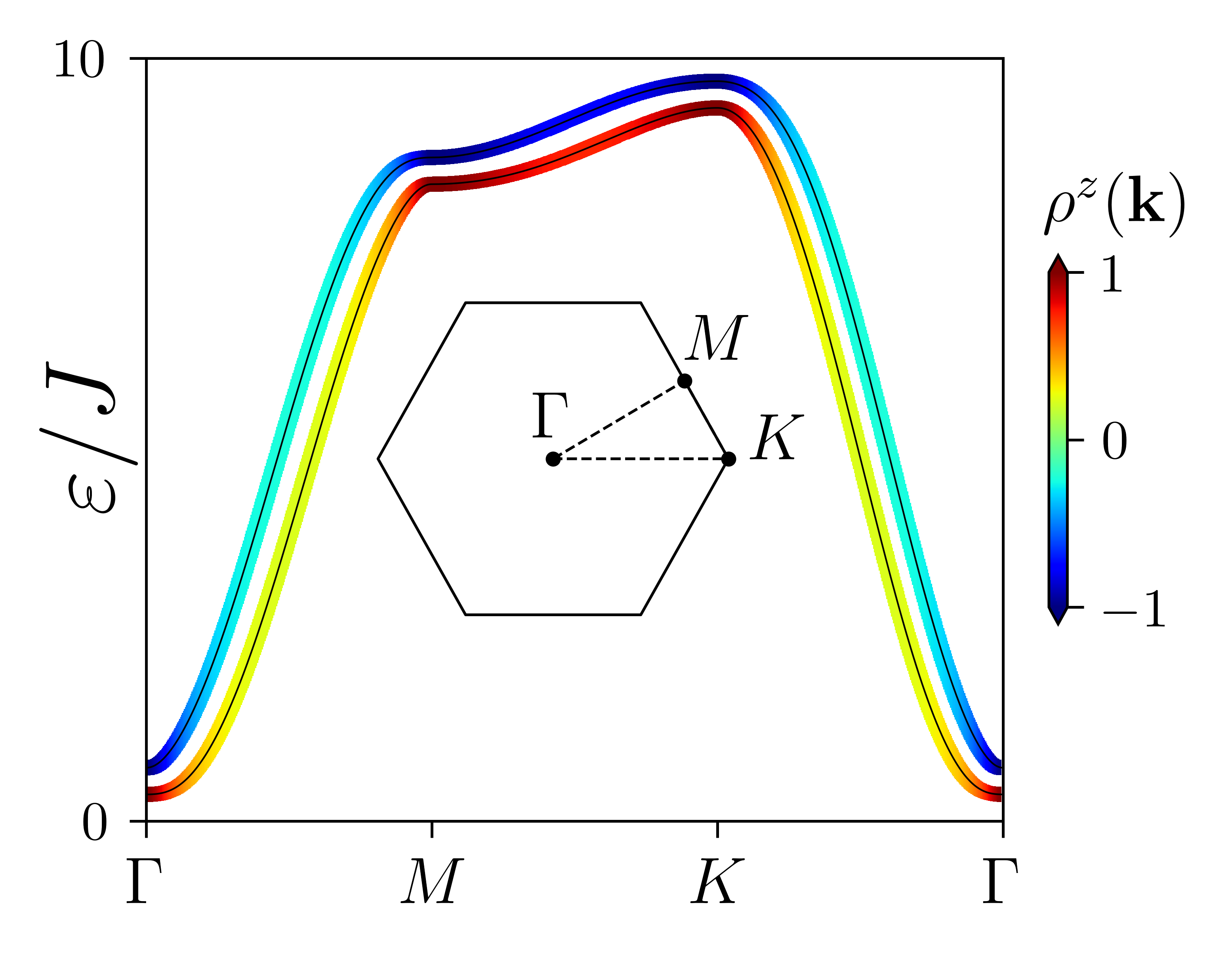}
\caption{Flavor-wave spectrum of the FM state in a triangular lattice with parameters: $J=1\,\mathrm{mV}$, $D=0.3J$, and $h=0.7J$. In color: the $z$-component of the color charge density. Observe that $\rho^{z}\to 1 (-1)$ for pure dipole (quadrupole) excitations.}
\label{fig:fm_band}
\end{figure}

The calculated band structure along $\Gamma$--$M$--$K$--$\Gamma$ is displayed in \Fig{fig:fm_band}. Its two branches reflect the enlarged local fluctuation space of a spin-1 moment. In the long-wavelength regime $|\bm k|\ll a^{-1}$, the reported limiting expressions are $|u^{(-)}(\bm k)|\approx1$, $|v^{(-)}(\bm k)|\approx0$, $\varepsilon^{(-)}(\bm k)\approx\frac32J(a k)^2+\frac12g\mu_BB$, and $|u^{(+)}(\bm k)|\approx0$, $|v^{(+)}(\bm k)|\approx1$, $\varepsilon^{(+)}(\bm k)\approx\frac32J(a k)^2+g\mu_BB$. Thus the normal-mode operators approach the dipolar and quadrupolar limits at small momenta, while finite mixing produces modes with both types of character.
\begin{figure}[t]
\centering\includegraphics[width=\columnwidth]{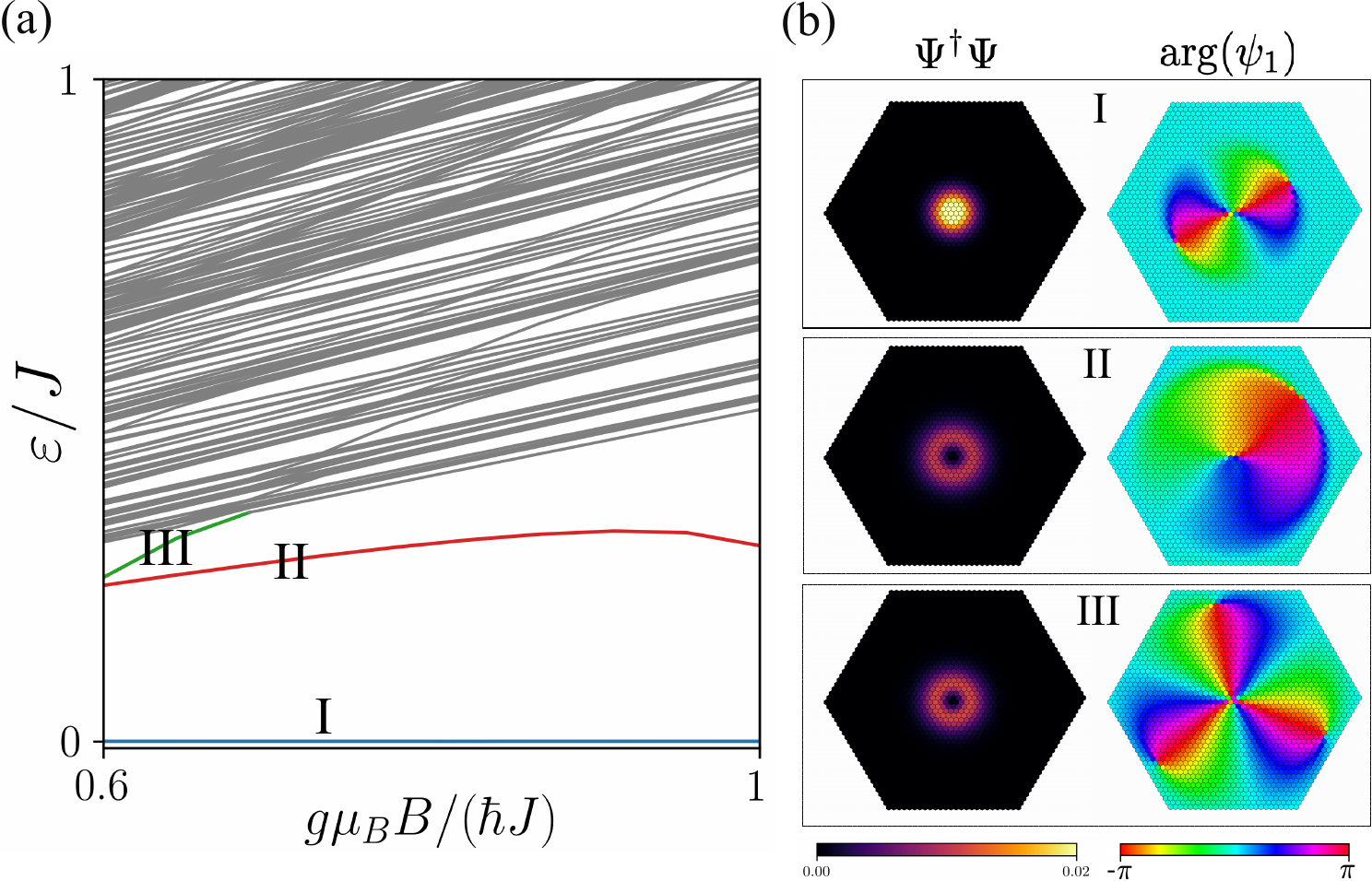}
\caption{Excitation spectrum of an isolated metastable $\CP$ skyrmion. (a) Eigenenergies as a function of the external magnetic field, with bound modes I, II, and III highlighted. (b) Spatial amplitudes and phases of the three bound modes, for the case $h=0.7 J$.}
\label{fig:bound_modes}
\end{figure} 

Similarly, we compute the spectrum of a isolated metastable skyrmion. \Fig{fig:bound_modes} displays the full spectrum as a function of the external magnetic field $h$. We found, in particular, localized modes below the extended excitation spectrum, labeled I, II, and III. Their different spatial amplitudes and phase patterns demonstrate that the texture supports distinct normal modes. In the present two-flavor description, these localized excitations can involve hybridzed dipole-quadrupole fluctuations, rather than being restricted to the single transverse field of conventional linear spin-wave theory. 

To characterize the internal degree of freedom more generally, we introduce the generators $\hat{\bm\tau}=(\frac12 \mathbb{I}_2,\frac12\bm\sigma)$ of $\mathfrak u(2)$. The corresponding color-charge density and current density are defined by
\begin{align}
\rho^a(\bm k)&=\Psi_{\bm k}^\dagger(\mathbb{I}_2\otimes\hat\tau_a)\Psi_{\bm k},\\
\mathcal J^a(\bm k)&=\Psi_{\bm k}^\dagger
(\Sigma_z\otimes\hat\tau_a)\hat{\bm v}\Psi_{\bm k},
\end{align}
where $\hat{\bm v}=i[\hat H,\hat{\bm r}]$ is the velocity operator. The integrated charges are $\mathcal{Q}^a=\sum_{\bm k}\rho^a(\bm k)$, which satisfy $[\mathcal{Q}^a,\mathcal{Q}^b]=i\epsilon^{abc}\mathcal{Q}^c$ in the $\mathfrak{su}(2)$ sector. The three-component vector $\bm{\mathcal{Q}}(t)=(\mathcal{Q}^1,\mathcal{Q}^2,\mathcal{Q}^3)^T$ is called the isospin. In the long-wavelength regime, the reported effective two-flavor Hamiltonian takes the form $\hat h(\bm k)\approx {B}\sigma_z/{2}+ {3J}k^2/2+
{D}\bm k\cdot\bm\sigma/{\sqrt2}$. We see that the last term can be interpreted as an isospin-orbit coupling. We next discuss how the nonuniform $\CP$ background modifies this two-component dynamics.

\section{Non-Abelian gauge theory of flavor-waves in $\CP$ textures}\label{sec:gauge}

The existence of two local fluctuation flavors makes it possible to describe their propagation using a matrix-valued gauge connection. In antiferromagnetic systems, analogous non-Abelian descriptions have been used to analyze the motion and polarization of magnon wavepackets \cite{Daniels2018,Daniels2019}; in the present model, the internal basis instead consists of the two tangent directions of a $\CP$ coherent state. We first formulate the continuum theory and then consider its consequences for scattering by an isolated skyrmion. This treatment complements the lattice Bogoliubov calculation and makes explicit how changes in the local order parameter can act on the isospin of a propagating excitation.

When the texture varies on length scales much larger than the lattice constant $a$, the continuum energy is written as an $\SU(3)$-gauged nonlinear sigma model \cite{Akagi2021_2},
\begin{equation}
\mathcal E(\bm n)=\frac12(\mathcal D_i\bm n)(\mathcal D^i\bm n)
-\bm B'\cdot\bm n.
\label{eq:continuum}
\end{equation}
Here $\mathcal D_i n^\mu=\partial_i n^\mu+\kappa f_{\mu\nu\eta}\Gamma_i^\nu n^\eta$, with $\kappa=D/(aJ)$, generalizes the helical derivative familiar from chiral magnets. For the interactions considered here, the corresponding background components are $\Gamma_x^\nu=\delta_{\nu7}$ and $\Gamma_y^\nu=-\delta_{\nu5}$. The effective field is determined by $\bm B'\cdot\bm n=\bm B\cdot\bm n/J-(\kappa[\bm\Gamma_i,\bm n])^2$, up to a constant; using the coherent-state constraint yields $B'_\mu=B_\mu/J-(\kappa^2/\sqrt3)\delta_{\mu8}$.

Now, expanding Eq.~\eqref{eq:continuum} to quadratic order in $\Psi=(\psi_1,\psi_2,\psi_1^\dagger,\psi_2^\dagger)^T$ about the equilibrium state produces the continuum flavor-wave Hamiltonian
\begin{equation}
H=\Psi^\dagger\hat{\mathcal H}\Psi
=\frac12(\nabla^A\Psi)^\dagger\cdot(\nabla^A\Psi)
+\Psi^\dagger\hat V\Psi,
\label{eq:nonabelian_hamiltonian}
\end{equation}
with the covariant derivative $\nabla^A\Psi=\nabla\Psi-i\sigma_z\hat A\Psi$. The connection and effective potential have matrix elements
\begin{align*}
[\hat A_j]_{\alpha\beta}&=i\bm e_\alpha\cdot\mathcal D_j\bm e_{-\beta},\\
[\hat V]_{\alpha\beta}&=\mathcal D_j\bm e_\alpha\cdot\mathcal D_j\bm e_{-\beta}
-(A_j)^\alpha{}_{\gamma}(A^j)^\gamma{}_{\beta}+\bm B\cdot T_{\alpha\beta}.
\end{align*}
Both $\hat A$ and $\hat V$ are Hermitian. The matrix character of the covariant derivative reflects the mixing between the two tangent-space components generated by spatial variations of the background.

\begin{figure}[t]
\centering\includegraphics[width=\columnwidth]{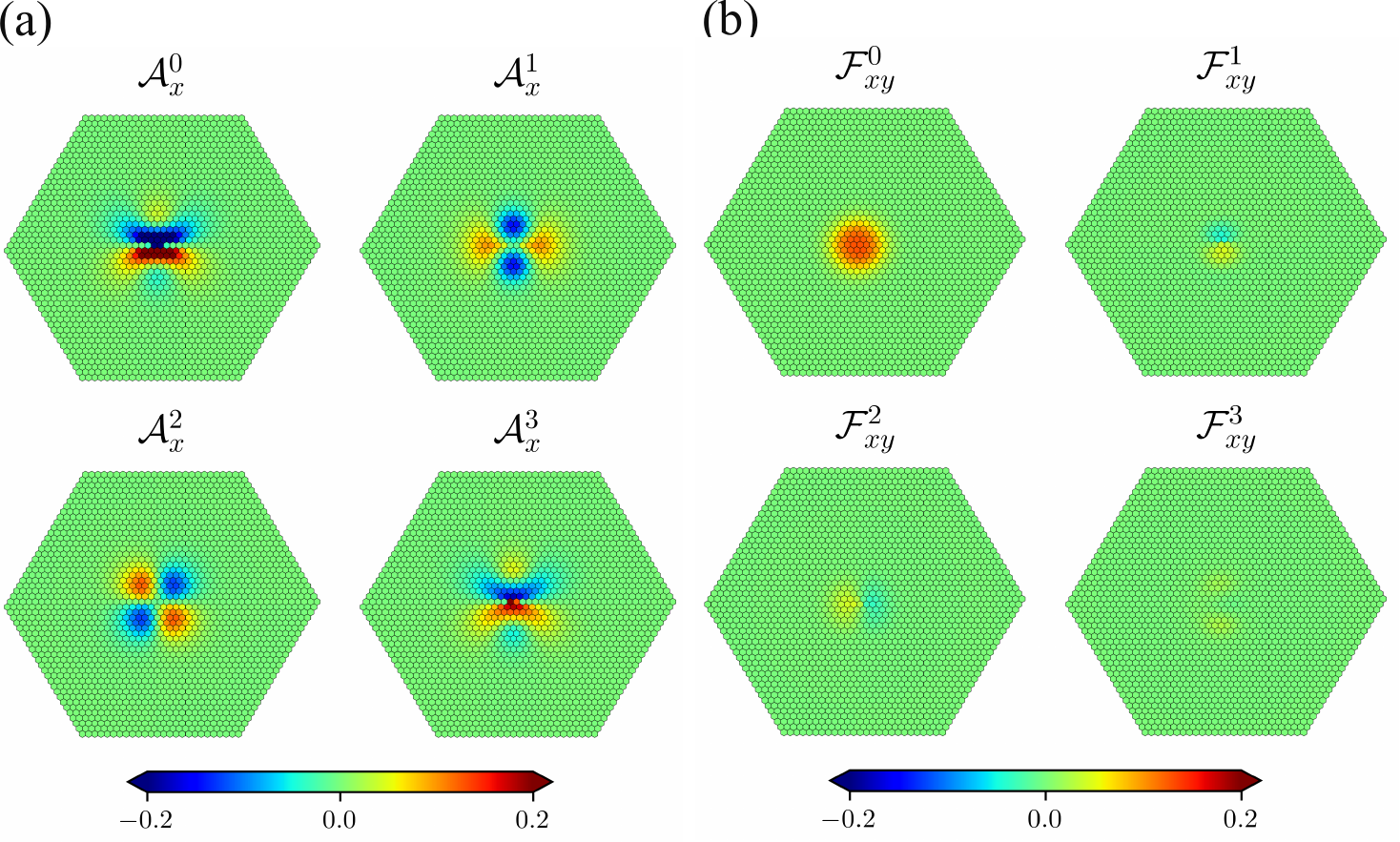}
\caption{(a) Components of the $\mathfrak{su}(2)$ connection $\hat A_{\mathrm{Sk}}=A_a^\mu\tau^a$ generated by the isolated skyrmion. (b) Corresponding components of the non-Abelian curvature $\hat{\mathcal F}_{xy}=\mathcal F_{xy}^a\tau^a$.}
\label{fig:gaugefield}
\end{figure}

To identify the relevant local frame structure, consider transformations $\hat U\in\SU(3)$ that leave the ground-state ray invariant: $\hat U|Z_{0}\rangle=e^{i\theta}|Z_{0}\rangle$. Their stabilizer is isomorphic to $U(2)$, and the associated tangent-frame freedom acts on the two-component flavor-wave field. We introduce $\tau^a=(\frac12 \mathbb{I}_2,\frac12\bm\sigma)$ and decompose the connection as $\hat A_i=\sum_{a=0}^3A_i^a\tau_a$. Its non-Abelian curvature is given in the original formulation by
\begin{equation*}
\hat{\mathcal F}_{ij}=\partial_i\hat A_j-\partial_i\hat A_j
-i[\hat A_i,\hat A_j].
\end{equation*}
Writing $\hat{\mathcal F}_{ij}=\epsilon_{ij}\mathcal F_{xy}^a\tau_a$, the matrix curvature makes explicit the possibility of flavor-dependent transverse forces and internal precession. The gauge field separates into the contribution associated with the spatially varying tangent frame and the term generated by the DM interaction:
\begin{equation*}
[\hat A_j]_{\alpha\beta}=
i\bm e_\alpha\cdot\partial_j\bm e_\beta^*
+\kappa f_{\mu\nu\eta}e_\alpha^\mu\Gamma_j^\nu e_\beta^{\eta*}.
\end{equation*}
The first term describes the texture-induced connection, while the second represents the chiral background. Far from the skyrmion core, the reported asymptotic form is $\hat A_j\approx{\kappa}\hat\tau_j/{\sqrt2}$, which can be removed at the order considered by the transformation $\widetilde\Psi(\bm r)=\exp[i\kappa\bm\sigma_z(\hat{\bm\tau}\cdot\bm r)/\sqrt2]\Psi(\bm r)$. Accordingly, the leading texture contribution is taken to be $[\hat A_j]_{\alpha\beta}=i\bm e_\alpha\cdot\partial_j\bm e_\beta^*$. The spatial distributions of the connection and curvature for the isolated skyrmion are displayed in \Fig{fig:gaugefield}.

We define the real-space color-charge density $\rho^a(\bm r)=\Psi^\dagger(\mathbb{I}_2\otimes\tau^a)\Psi$ and the corresponding current $\bm{\mathcal J}^a(\bm r)=-i\Psi^\dagger(\sigma_z\otimes\tau^a)\nabla^A\Psi$. Their evolution is described by the covariant continuity equation
\begin{equation*}
\partial_t\rho^a+\nabla^A\cdot\bm{\mathcal J}^{a}=s^a.
\end{equation*}
The source term is $s^a=i\Psi^\dagger([\hat{\mathcal H},\sigma_z]\tau^a)\Psi$. Thus, we find that the total color charges $\mathcal{Q}^{a}=\int \rho^a(\bm{r}) d^{2}r$ are not conserved quantities. This non-conservation effect is explained by the rotation of the isospin caused by the non-abelian emergent gauge field due to the $\CP$ texture.

\subsection{Flavor Hall effect of wavepackets by $\CP$ skyrmions}\label{sec:scattering}

We consider a normalized flavor-wavepacket $|\Psi_c\rangle$ incident on an isolated $\CP$ skyrmion. Its average position, momentum, and isospin are denoted by $\bm r_c(t)$, $\bm p_c(t)$, and $\bm{\mathcal{Q}}(t)$, respectively. Now we employ the semiclassical dynamics approach \cite{Chang1995,Sundaram1999}, taking into account for the non-abelian isospin dynamics \cite{Daniels2018,Daniels2019}. Using \ref{eq:nonabelian_hamiltonian}, we obtain that the Hamiltonian of the wavepacket becomes
\begin{equation*}
H(\bm r_c,\bm p_c,\bm{\mathcal{Q}})\approx
\frac{1}{2m}\bigl(\bm p_c-\mathcal{Q}_a\bm A^a(\bm r_c)\bigr)^2
+\mathcal V(\bm r_c),
\end{equation*}
where $m=(3Ja^2)^{-1}$ is the effective wavepacket mass. Using the isospin algebra $[\mathcal{Q}_a,\mathcal{Q}_b]=\epsilon_{abc}\mathcal{Q}_c$, the semiclassical dynamics is governed by
\begin{align*}
m\ddot{\bm r}_c&=-\nabla_{\bm r}\mathcal V(\bm r_c)
+\mathcal{Q}^a\mathcal F_{xy}^{a}(\bm r_c)\,\hat{\bm z}\times\dot{\bm r}_c,\\
\dot{\bm{\mathcal{Q}}}&=-[\dot{\bm r}_c\cdot\hat{\bm A}(\bm r_c),\bm{\mathcal{Q}}].
\end{align*}
The first equation describes the transverse Lorentz-like force arising from the emergent gauge field, while the second line is known as the Wong's equation \cite{Wong1970} and describes the parallel transport on the internal states, here the isospin space, along the trajectory of the wavepacket. 

In the short-wavelength limit $k \gg R_{\mathrm{sk}}^{-1}$, with $R_{\mathrm{sk}}$ being the radius of the skyrmion, the transverse deflection can be estimated by taking the unperturbed path to be a straight line $\bm r_0(t)=\bm v_0t$ and neglecting the scalar potential term. The resulting transverse velocity change is
\begin{equation*}
\Delta\bm v^{\perp}\approx\frac1m\int_{-\infty}^{\infty}\dd t\,
Q^0\mathcal F_{xy}^{0}(\bm v_0t)\hat{\bm z}\times\bm v_0
\approx\frac{\hat{\bm z}\times\hat{\bm v}_0}{\sqrt{\pi mR_{\mathrm{sk}}}}.
\end{equation*}
The estimate uses that $|F_{xy}^{0}|\gg |F_{xy}^{i=1,2,3}|$ (see \ref{fig:gaugefield}) and we approximate by a Gaussian profile $\mathcal F_{xy}^{0}(r)\approx\exp[-r^2/(2R_{\mathrm{sk}}^2)]/\sqrt\pi R_{\mathrm{sk}}$. Thus, we arrive at the Hall angle for the favor-wavepackets $\Theta_H=\tan^{-1}(\Delta v^{\perp}/v_0)\approx1/(\sqrt{\pi mR_{\mathrm{sk}}}v_0)$. 

Now, let us calculate the change of the isospin. The parallel transport along the trajectory produces a rotation of the isospin. Integrating the isospin equation yields the non-Abelian phase factor
\begin{equation*}
\exp(i\bm\Omega\cdot\hat{\bm\tau})
=\mathcal P\exp\left[\int_{\bm r_c}\hat{\bm A}_{\mathrm{Sk}}(\bm r)\cdot\dd\bm r\right],
\end{equation*}
where $\mathcal P$ denotes path ordering. Thus the outgoing isospin is $\bm{\mathcal{Q}}(t\to\infty)=\hat R(\bm\Omega)\bm{\mathcal{Q}}(t\to-\infty)$, where $\hat R(\bm\Omega)\in SO(3)$ is the associated rotation. Expressed in the laboratory frame, the reported transformation is $(\bm{\mathcal{Q}}_{\mathrm{Lab}}\cdot\bm\tau)=e^{i\kappa(\hat{\bm\tau}\cdot\bm r)/\sqrt2}(\bm{\mathcal{Q}}\cdot\bm\tau)e^{-i\kappa(\hat{\bm\tau}\cdot\bm r)/\sqrt2}$ and $\bm{\mathcal{Q}}_{\mathrm{Lab}}(t)=\hat R(\kappa\bm r_c(t)/\sqrt2)\bm{\mathcal{Q}}(t)$. These relations describe the combined background precession and the skyrmion-induced isospin rotation.

\begin{figure}[t]
\centering
\includegraphics[width=\columnwidth]{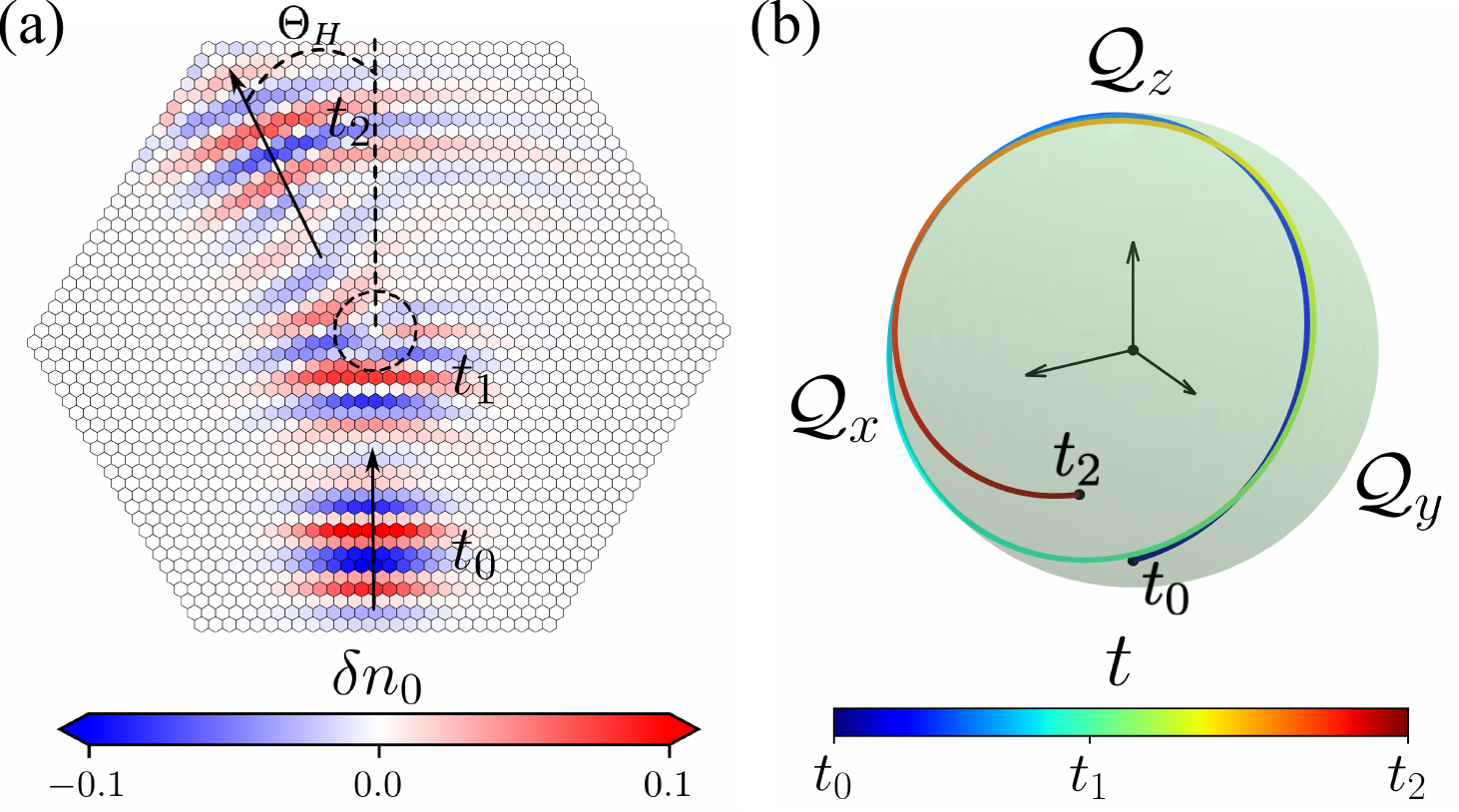}
\caption{Scattering of a flavor-wavepacket by an isolated $\CP$ skyrmion. (a) Fluctuation of the quadrupolar component $n_0$. (b) Time evolution of the wavepacket's isospin $\bm{Q}$ (normalized to the unit sphere), showing the precession effect during the scattering event.}
\label{fig:isospin}
\end{figure}

To verify our predictions, we performed numerical simulations of a flavor-wavepacket interacting with a $\CP$ skyrmion by solving the generalized Landau-Lifshitz-Gilbert equation. The results are displayed in Fig. 2. We find a left-ward scattering of the flavor-wavepacket from the skyrmion, in agreement with the theorical prediction. Similarly, we find Hall effect for coherent flavor-wavefront upon the skyrmion texture (see Fig. 1 in the Supplemental Material). 

The wavepacket evolution in \Fig{fig:isospin} illustrates the simultaneous influence of the texture on spatial propagation and the internal flavor degree of freedom. A transverse deflection and an isospin rotation are different aspects of the same gauge-field description: the former follows from the curvature sampled along the trajectory, while the latter reflects the accumulated matrix-valued connection. This behavior distinguishes the two-flavor problem from the single-component description of magnons in an ordinary dipolar background. It also motivates the analysis of transverse transport in a periodic $\CP$ skyrmion configuration.

\section{Topological modes in $\CP$ skyrmion crystals}\label{sec:crystal}

\begin{figure}[t]
\centering\includegraphics[width=\columnwidth]{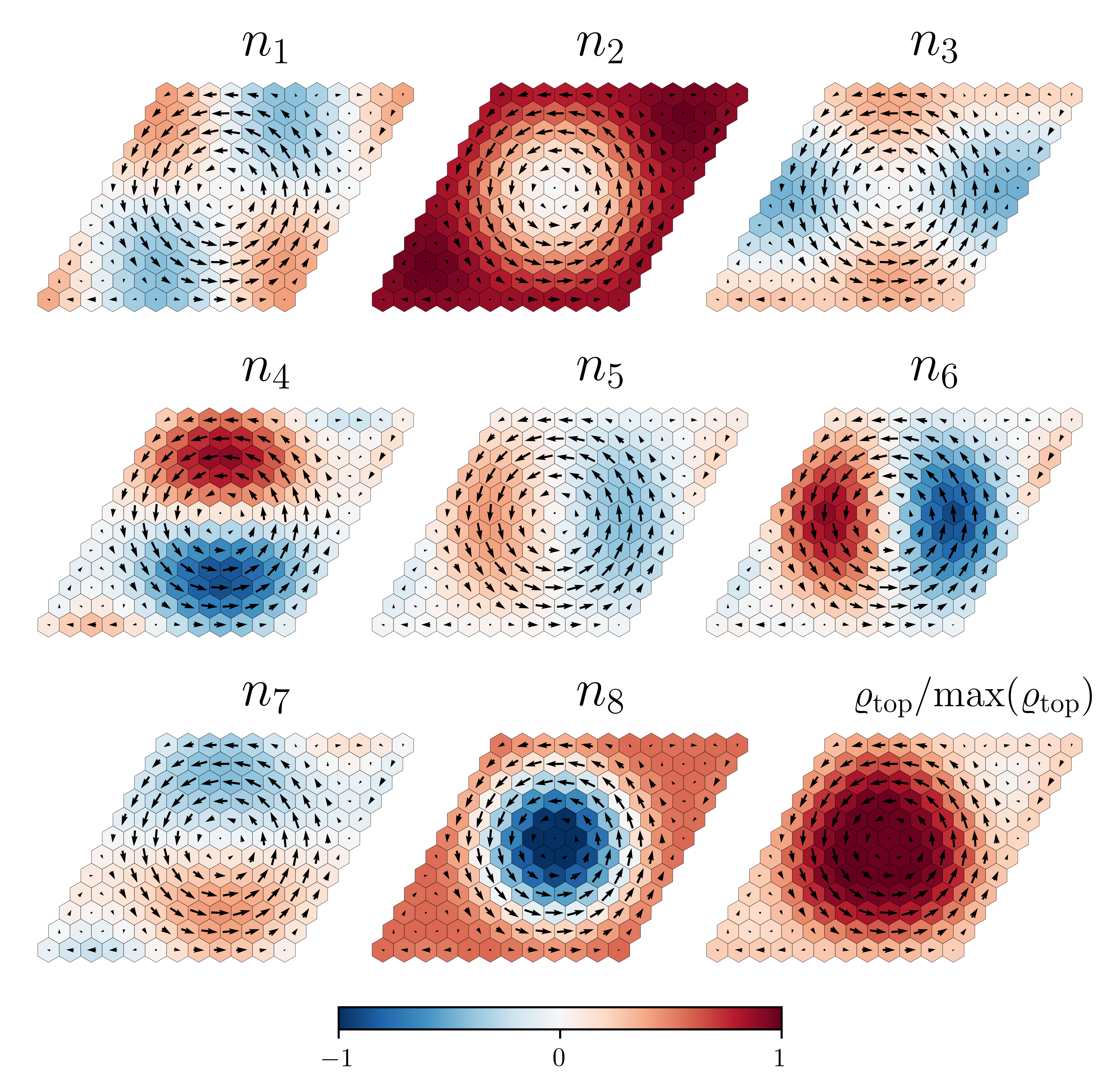}
\caption{Numerically relaxed $\CP$ skyrmion crystal on the triangular lattice. The panels show the components $\{n_\mu\}_{\mu=1,\ldots,8}$ and the three-dimensional dipole moment $\langle\bm S\rangle=(n_7,-n_5,n_2)^T$. The resulting Bloch-type texture has $\mathcal{Q}_{\mathrm{top}}\approx 1$ and a unit cell of $L\times L$ sites with $L=12$. The magnetic parameters used are $J=1\ \mathrm{mV}$, $D=0.3J$ and $g\mu_B B=0.4J$.}
\label{fig:crystal}
\end{figure}

We now turn to the flavor-wave spectrum of a periodic $\CP$ skyrmion texture. The ground state is determined through Monte Carlo annealing \cite{Zhang2023,Dahlbom2022,Wenzel2013,Wang2012SpinDO,Remund2022} followed by a gradient descent according to Eq.~\eqref{eq:llg} until to reach the ground state \cite{Zhang2023,Dahlbom2022,Wenzel2013,Wang2012SpinDO,Remund2022}. The magnetic unit cell is represented by $L\times L$ atomic sites with $L=12$, the magnetic parameters are $J=1\ \mathrm{mV}$, $D=0.3J$, while we vary $h=g\mu_B B$ from $0$ to $1\ \mathrm{mV}$. The ground state configurations are obtained are clasified as \cite{Amari2022,Zhang2023}: skyrmion crystal of first kind (SkX-I), consisting in a honeycomb lattice of half-skyrmions; skyrmion crystal of second kind (SkX-II), being a triangular lattice of skyrmions; and the fully polarized ferromagnetic state (FM). In particular, Fig. \Fig{fig:crystal} shows the obtained periodic ground state for $h=0.4J$. The computed topological charge, obtained by numerical evaluation of Eq.~\eqref{eq:charge}, is $\mathcal{Q}_{\mathrm{top}}\approx 1$. This value is close to the unit charge expected for the corresponding continuum configuration. The resulting texture is identified in the original calculation as a Bloch-type $\CP$ skyrmion.

\begin{figure}[t]
\centering\includegraphics[width=\columnwidth]{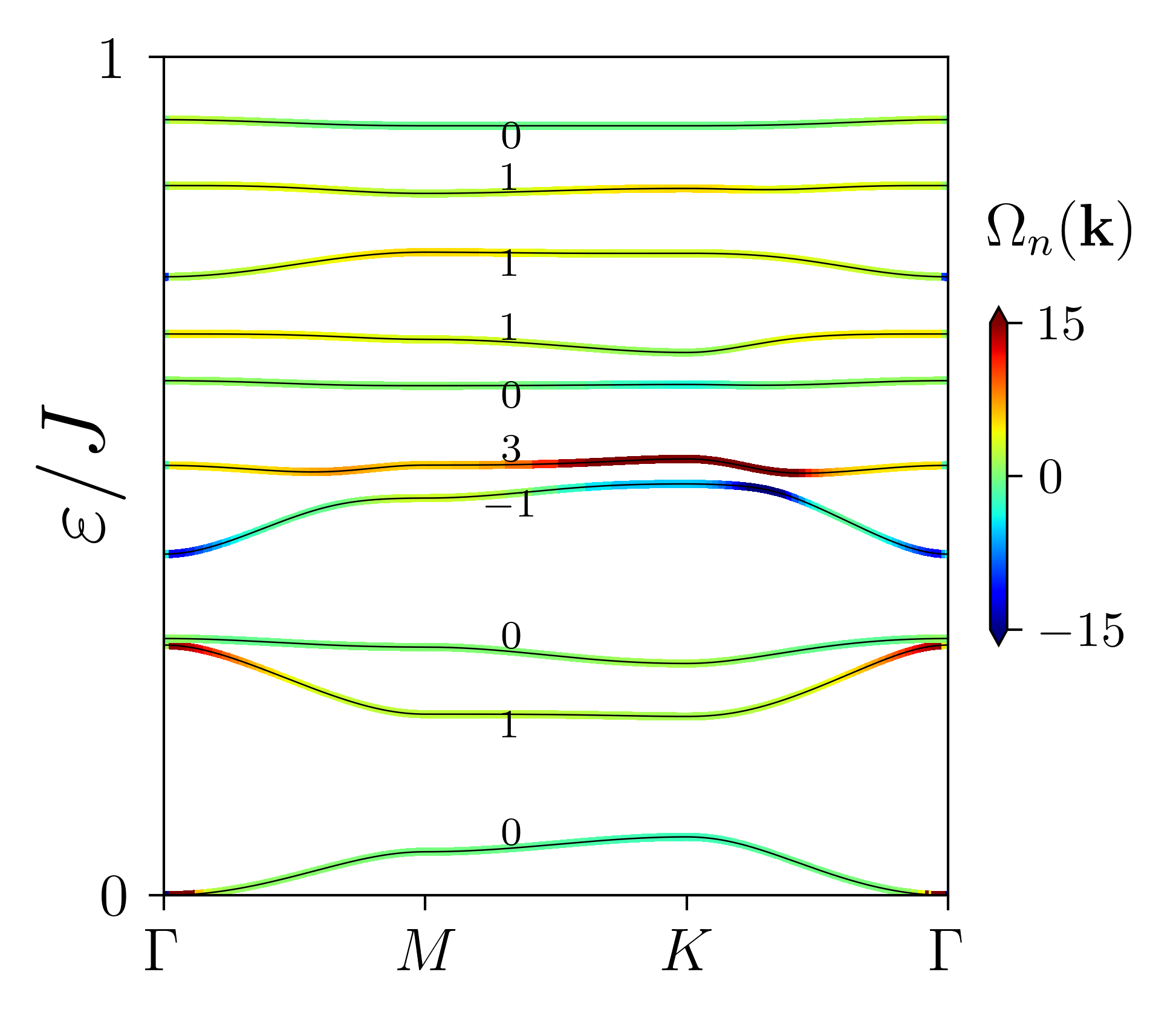}
\caption{Flavor-wave band structure of the $\CP$ skyrmion crystal depicted at Fig. \Fig{fig:crystal}. In color: the Berry curvature $\Omega_{n}(\bm k)$, with the respective Chern number $C_{n}$ highlighted, of the first $10$ flavor-wave bands.}
\label{fig:chernbands}
\end{figure}

The excitation spectrum is obtained by applying the quadratic bosonic Hamiltonian derived in Sec.~\ref{sec:flavor} to this periodic background. Bloch periodicity allows the two-flavor Bogoliubov problem to be solved throughout the Brillouin zone. The resulting bands, shown in \Fig{fig:chernbands}, display the collective modes of the skyrmion crystal, including the coupling between dipolar and quadrupolar fluctuations induced by its spatially varying order parameter. In contrast to an isolated texture, the periodic configuration supports dispersive bands whose geometry can be characterized in momentum space. We introduce the projector on the $n$-th bosonic band as $\mathcal{P}_n(\bm k)=U(\bm k)\Pi_n\Sigma_zU^\dagger(\bm k)\Sigma_z$, where $[\Pi_n]_{ij}=\delta_{in}\delta_{jn}$. The Berry curvature is evaluated through
\begin{equation*}
\Omega_n(\bm k)=i\epsilon_{ij}\operatorname{Tr}
\left[\mathcal{P}_n(\bm k)\partial_{k_i}\mathcal{P}_n(\bm k)\partial_{k_j}\mathcal{P}_n(\bm k)\right],
\end{equation*}
and the corresponding Chern number is
\begin{equation}
C_n=\frac{1}{2\pi}\int_{\BZ}\Omega_n(\bm k)\,\dd^2k.
\label{eq:chern}
\end{equation}
For the first bands, the reported numerical values of the Chern numbers and the Berry-curvature distribution is displayed in \Fig{fig:chernbands}.
The distinction between real-space and band topology is important here. The quantity $\mathcal{Q}_{\mathrm{top}}$ characterizes the winding of the static spinor texture, whereas $C_n$ characterizes the geometry of an individual dynamical eigenband. Although the same nonuniform order influences both, their definitions involve different spaces and need not yield identical integers. Consequently, the observation of a nearly unit skyrmion charge is not by itself sufficient to infer the Chern numbers; the latter follow from the explicit eigenmode calculation.

\subsection{Thermal flavor Hall effect in $\CP$ skyrmion crystals}

In recent years, thermal Hall effect in antiferromagnetic skyrmion lattice via non-abelian gauge symmetry has been studied in \cite{Takeda2024,Kawano2025}. On the other hand, Nernst effects of linear flavor-waves about the quadrupolar order state is proposed in \cite{Lu2024}. Therby, it becomes a relevant case of study the emergence of flavor thermal Hall effect in $\CP$ skyrmion crystals. 

The momentum-space Berry curvature of stable bosonic bands contributes to the intrinsic thermal Hall conductivity. In the standard noninteracting quasiparticle approximation, including the conventional energy-magnetization correction, the result is \cite{Matsumoto2011}
the thermal Hall conductivity
\begin{equation}\label{eq:kappa}
\kappa_{xy}=-\frac{k_B^2T}{\hbar}
\sum_n\int_{\BZ}\frac{\dd^2k}{(2\pi)^2}
 c_2\!\left[n_B(\varepsilon_n(\bm k))\right]\Omega_n(\bm k).
\end{equation}
Here $c_2(\rho)=(1+\rho)\left[\log\frac{1+\rho}{\rho}\right]^2-(\log\rho)^2-2\operatorname{Li}_2(-\rho)$, with $\operatorname{Li}_2$ the dilogarithm, and $n_B(\bm k)=(e^{\hbar\omega_n(\bm k)/(k_BT)}-1)^{-1}$ denotes the Bose--Einstein distribution. The sum is over all positive-frequency BdG branches retained in the approximation. If only a subset of bands is included, the plotted result must be labeled a partial-band contribution, not the total conductivity. The response depends on Berry curvature weighted by thermal occupation and is not determined by the Chern numbers alone. Bands with opposite curvature may partially cancel, and the sign of $\kappa_{xy}$ can change with temperature or field without any single-band Chern number changing.

\begin{figure}[t]
\centering\includegraphics[width=0.90\columnwidth]{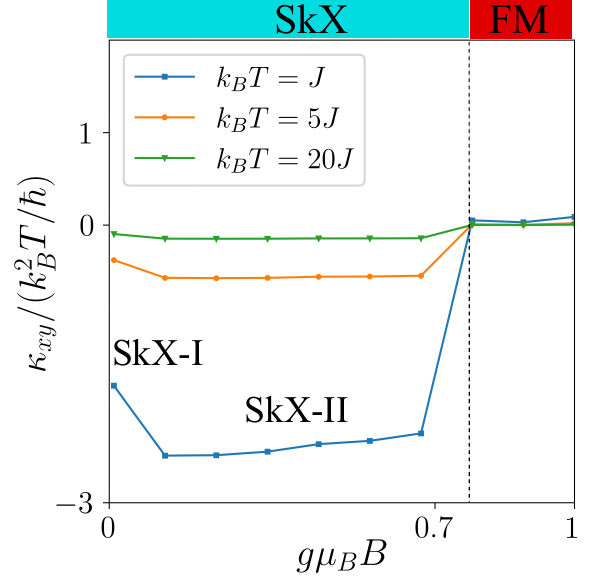}
\caption{Thermal Hall conductivity of the $\CP$ ground state as a function of the external magnetic field for diferrent temperatures. This reflects of the flavor-wave bands topology of the skyrmion phase SkX.}
\label{fig:thermal}
\end{figure}

The calculated field and temperature dependence is presented in \Fig{fig:thermal}. The finite transverse response is consistent with the nonzero Berry curvature found in the positive-energy collective bands. The most pronounced low-temperature structure occurs in a limited field range, while the higher-temperature curves display a different magnitude and can approach cancellation. These trends should be interpreted as changes in the weighted sum of band curvatures, rather than as a direct plot of the Chern numbers. In particular, curves at $k_BT$ substantially larger than $J$ illustrate the formal occupation-weighted band formula, but should not be interpreted as quantitatively reliable finite-temperature predictions without verifying the survival of the skyrmion crystal and the applicability of the harmonic approximation. Any sharp field dependence could result from changes in gaps, redistribution of Berry curvature, or changes of the equilibrium texture; assigning a topological phase transition would require an explicit calculation of the relevant gaps and Chern numbers as functions of field.

The hybrid nature of the modes also raises the prospect of spin- and quadrupole-resolved transverse responses. Such coefficients cannot, however, be obtained in general by multiplying Eq.~\eqref{eq:kappa} by an arbitrarily defined local flavor weight. With DM interactions the spin projection need not be conserved, and quadrupolar density is not a generic conserved transport charge. A well-defined spin or quadrupolar Nernst coefficient requires a specified current operator, its torque/source terms, and an appropriate linear-response calculation \cite{Lu2024}. The thermal Hall conductivity is the most direct band-geometric transport observable within the present calculation.

\section{Discussion}\label{sec:discussion}

Our findings connect three distinct layers of geometry. First, the spin-one coherent-state manifold $\CP$ permits topological configurations involving both dipolar and quadrupolar degrees of freedom. Second, fluctuations around such configurations form a complex rank-two tangent space, and their propagation is described naturally by a $U(2)$ connection. Third, the periodic crystal supports bosonic bands whose Berry curvatures and Chern numbers characterize topology in momentum space. Although all three structures originate from the same multipolar background, they are neither identical nor mutually sufficient: a nonzero real-space charge does not fix individual band Chern numbers, and a non-Abelian local frame description does not automatically imply a thermal Hall effect.

The results also specify a distinction from existing non-Abelian antiferromagnetic magnonics. In antiferromagnetic skyrmion lattices, multiple internal components can arise from sublattice-dependent spin-wave polarizations, leading to effective $SU(3)$ transport and thermal Hall phenomena \cite{Takeda2024,Kawano2025}. Here the two flavors already exist in the local Hilbert space of an individual spin-one moment and are associated with different dipolar--quadrupolar transition channels. This is a new source of multicomponent wave dynamics even for a lattice with one magnetic site per primitive unit cell. Nevertheless, comparison with the antiferromagnetic case requires physical observables, since the dimension of a local gauge representation alone cannot distinguish scattering mechanisms.

A second consequence is the possibility of detecting the same collective band in complementary probes. Ordinary magnetic neutron scattering primarily accesses correlations of dipole operators, whereas quadrupole-sensitive correlations may require polarization analysis, nonlinear spectroscopy, or coupling to higher-order spin operators \cite{Bai2021,Watanabe2025}. More direct evidence for the $SU(2)$ part of the connection would come from preparing different incident flavor polarizations and comparing outgoing amplitudes or from measuring noncommuting flavor rotations along different controlled paths. The thermal Hall signal, in contrast, measures the combined occupied-band geometry and need not isolate the traceless gauge structure by itself.

The present analysis has identifiable limitations. It is based on a coherent-state background and a harmonic two-flavor expansion; quantum corrections, magnon--magnon interactions, damping, and finite-temperature texture fluctuations may renormalize the bands and their transport coefficients \cite{Koyama2023,Remund2022}. In addition, the existence of nonzero Chern numbers suggests chiral boundary modes in isolated bulk gaps, but an explicit ribbon calculation is needed to establish their dispersion, polarization, and localization \cite{RoldanMolina2016,Shindou2013}. A quantitative comparison to experiment also requires material-specific coupling constants and a controlled determination of the stable $\CP$ crystal regime. These limitations do not alter the key geometric conclusion that the local fluctuation space of a spin-one projective texture supports two coherently coupled flavors.

\section{Conclusions}\label{sec:conclusions}

We have developed a coherent-state flavor-wave framework for dipolar and quadrupolar excitations propagating on $\CP$ skyrmion backgrounds. The two complex tangent directions of the spin-one order parameter give rise to a bosonic BdG description with two local excitation flavors. In a polarized reference state they are naturally identified with one- and two-unit spin-flip transitions, whereas the nonuniform projective texture mixes their character through position-dependent intersite couplings.

The local rotation of this two-dimensional tangent frame defines a matrix-valued $U(2)$ connection. Its curvature and Wilson lines provide a geometric language for wavepacket deflection and internal isospin evolution, with the important qualification that $U(2)$ is a local frame freedom and that physical non-Abelian signatures must be distinguished from the Abelian component of the response. Numerical results for isolated skyrmions exhibit localized excitation modes and asymmetric scattering, while a relaxed periodic $\CP$ skyrmion texture supports bands with finite bosonic Berry curvature and nonzero Chern numbers. Their thermal occupation leads to a transverse thermal response, whose temperature and field dependence reflects contributions from the full positive-energy spectrum rather than the Chern numbers alone. These results establish a connection between multipolar magnetic textures and topological flavor-wave transport, extending skyrmion magnonics beyond single-dipole excitations. Further work should focus on converged band-resolved topological indices, explicit chiral edge spectra, polarization-resolved scattering, and the stability of the proposed responses under interactions and realistic experimental conditions.

\begin{acknowledgments}
C.S. acknowledges support from Proyecto CORFO/ANID Science UP "Impulsando capacidades de innovación, emprendimiento y transferencia tecnológica para el desarrollo del país", Código 20CEIN2-142145.
E.S. acknowledges partial financial support from DICYT Regular Grant No. 042631SD\_REG.
Funding is acknowledged from Fondecyt Regular 1230747 and 1230515, and ANID CEDENNA CIA 250002 
\end{acknowledgments}

\bibliography{Biblio}

\end{document}